# Identification of Potential Replacement Materials for Lead in $CH_3NH_3PbI_3$ using First Principle Calculations

Kamal Choudhary

Department of Materials Science and Engineering, University of Florida, Gainesville, FL, 32611

**Abstract**: There is a need for replacement material for lead due to its toxicity in organic-inorganic hybrid halide $CH_3NH_3PbI_3$ to facilitate its wide range photovoltaic applications. Here, first-principle calculations are used to screen some of the elements from the periodic table based on the band-gap, Goldschmidt's tolerance factor, absorption coefficient and carrier effective mass obtained in the resultant material. The oxidation state of the replacement material are emphasized and their valence band maxima (VBM)- conduction band minima (CBM) alignment are compared to $CH_3NH_3PbI_3$ during the selection procedure.

Methylamine lead iodide ($CH_3NH_3PbI_3$ or $MAPbI_3$) has revolutionized the area of photovoltaic applications due to its high efficiency and low cost compared to conventional inorganic material based technology [1]. However, the toxicity of lead hinders the commercial utilization of the material at large scale. Lead poisoning can cause a variety of diseases to various organs and tissues, including the heart, bones, intestines, kidneys, and reproductive and nervous systems [2]. Recently, tin (Sn) was proposed as the non-toxic replacement of Pb, but it suffers from the instability issue due to its existence in multiple oxidation state; +2 and +4 [3,4]. Hence, a high throughput screening of material based on a reliable and computationally efficient method such as density functional theory (DFT) [5] is necessary to guide the experimental exploration for solving the problem. As mentioned above, it is necessary for the replacement material to remain stable in a +2 oxidation state in its halide form as the hybrid material is generally prepared via solution route using the halide of the material and methylamine [6,7]. Hence, lists of elements from the periodic table with +2 oxidation state were chosen for DFT calculations. Important criteria used during screening of the material was the band-gap ($E_g$) of the material satisfying Shockley and Queisser (SQ) [8] ($E_g \sim$ 1.7 eV to 3.3 eV), Goldschmidt's tolerance [9] factor ($t = \frac{r_{MA} + r_I}{\sqrt{2}(r_X + r_I)}$, for $MAXI_3$, r represents ionic radii), absorption coefficient (used in calculation for maximum solar efficiency [10]), and carrier effective mass [11] (for electron, $m_e$; for hole, $m_h$).

Although originally introduced by Weber [12], Kojima et al. [13] introduced $MAPbI_3$ for solar cell applications in 2009, there is still much to be done in terms of DFT study for the hybrid material such as $MAPbI_3$. To list a few of them, Mosconi e al.[14] studied the structural and optical properties of $CH_3NH_3PbX_3$ and $CH_3NH_3PbI_2X$ perovskites (X = Cl, Br, I). They showed

DFT can successfully be used to calculate both lattice constant and band-gap of the material, comparable to experiments. Jishi et al. [15] compared the structural and electronic properties of $CH_3NH_3PbI_3$, $CH_3NH_3PbBr_3$, $CsPbX_3$ (X=Cl, Br, I), and rubidium lead iodide ($RbPbI_3$) using TB-mBJ exchange-potential. Feng et al. [16] studied the tetragonal and orthorhombic structures of the material and showed the effective masses were anisotropic in nature and also calculated the theoretical absorption spectra of the material. Furthermore, they showed that van der Waals interactions were important for obtaining the accurate equilibrium cell volumes from DFT. In a later work, Feng et al. [17] showed that using Sn as substituent of Pb, absorption efficiency could be increased due to reduction of bandgap of the material and they confirmed that the orthorhombic representation [18] for the hybrid material is justifiable over tetragonal and cubic structures. Amat et al. [9] showed the importance of Goldshmidt factor for the hybrid material and showed its importance in determining structural stability using first-principle calculations. Umari et al. [19] studied the effect of various levels of DFT (spin-orbit coupling, GW-methods) on the electronic properties such as band-gap and effective mass for $CH_3NH_3$ (Sn, Pb) $I_3$ to confirm the necessity of these high-level calculations . Haruyama et al. [20] showed the effect of surface terminations on stability and electronic properties of $MAPbI_3$. They showed flat terminations under the $PbI_2$ rich condition were advantageous in regards to solar cell performance. While there are so many ongoing works for optimizing the stability and efficiency of the hybrid pervoskite materials, not much has been done for lead replacement elements options.

It is to be noted that the lead iodide crystallizes into rhombohedral, P-3m1 space group. Hence, iodides with same space group with band-gap in the visible region, appropriate Goldschmidt's factor, absorption coefficient comparable to $MAPbI_3$ within visible range and low effective mass of carriers should be preferred over others as potential replacement candidates. Goldschmidt factor is an important factor in studying the stability of pervoskite materials of type $ABX_3$. For a stable pervoskite the factor is closer to 1.0. However, for the present case of $MAXI_3$ (X = replacing element of Pb), MA is a molecular specie instead of a monoatomic ion. Hence, the ionic radius of MA is controversial. Amat et al. [9] calculated the effective ionic radius of the molecule based on DFT to be 2.7 Å for $MAPbI_3$, while Yin et al. calculated it to be 2.37 Å [10]. Here, we take the data by Amat et al. [9] for the calculation of Goldschmidt's factor. Based on this data, and the ionic radii of the divalent substituent ion, the factor was calculated and shown in Table. 1

All calculations were carried out using Vienna Ab initio Simulation Package (VASP) package [21,22]. An structure consisting of 48 atoms was taken and its complete structural optimization was performed with a 4x4x4 k-mesh using Perdew, Burke and Ernzerhof (PBE) functional [23] and with van der Waal's interaction (PBE+D2) [24] separately. A 2x1x2 supercell for the structure is shown in Fig. 1. The projector-augmented wave (PAW) approach, developed by Bloch in VASP was used for the description of the electronic wavefunctions. Plane waves have been included up to an energy cut-off 600 eV. The energy criterion convergence was $10^{-6}$ eV. Grimme's D2 parameters [24] were taken for including van der Waal's interactions. Scalar relativistic effects were taken into account in the pseudopotential. After the geometric optimization, hybrid functional to Heyd-Scuseria-Ernzerhof (HSE06) based electronic optimization was carried out on the geometrically optimized structure from D2 correction at gamma point only. Hybrid functional is employed because it is proven to give better understanding of the electronic properties of material [25,26] at much reasonable computational cost. Hybrid functional mix 25 % of exact nonlocal exchange of Hartree-Fock theory [27] with the density functional exchange. Subsequently, effect of spin-orbit coupling was also monitored along with hybrid functional. The effect of spin-orbit coupling [28] becomes necessary especially for the heavy atoms such as transition elements and group-IV elements. Absorption coefficient was obtained from calculating dielectric function for materials, which is obtained as:

$$\varepsilon_{\alpha\beta}^2(\omega) = \frac{4\pi^2 e^2}{\Omega} \lim_{q\to 0} \frac{1}{q^2} \sum_{c,v,k} 2w_k \delta(\varepsilon_{ck} - \varepsilon_{vk} - \omega) \times \langle u_{ck+e_\alpha q} | u_{vk} \rangle \langle u_{ck+e_\beta q} | u_{vk} \rangle^* \qquad (1)$$

Here the indices $c$ and $v$ refer to conduction and valence band states respectively, $u_{ck}$ and $w_k$ are the cell periodic part of the orbitals and weight of k-point respectively. $\Omega$ is cell volume and $e$ is electronic charge. The effective mass of the hole and electron are obtained by fitting parabolic expression eq. 2 around gamma point in Brillouin zone, diagonalizing the matrix A then taking inverse of the matrix as in eq. 3 at the band extremes.

$$E(\vec{k}) = E(\vec{k}_\Gamma) \pm \frac{\hbar^2}{2} [\vec{k}_i - \vec{k}_\Gamma] A_{ij} [\vec{k}_i - \vec{k}_\Gamma]^T \qquad (2)$$

where $E(\vec{k}_\Gamma)$ represents the eigenenergies at the band extremes, that is, the values at the gamma point, $\vec{k}_\Gamma$ in the Brillouin zone. The tensor $A_{ij}$ is related to the effective mass tensor $m_{ij}^*$ by

$$[A_{ij}] = \sum_{i=1}^{3} \sum_{j=1}^{3} \frac{1}{m_{ij}^*} \qquad (3)$$

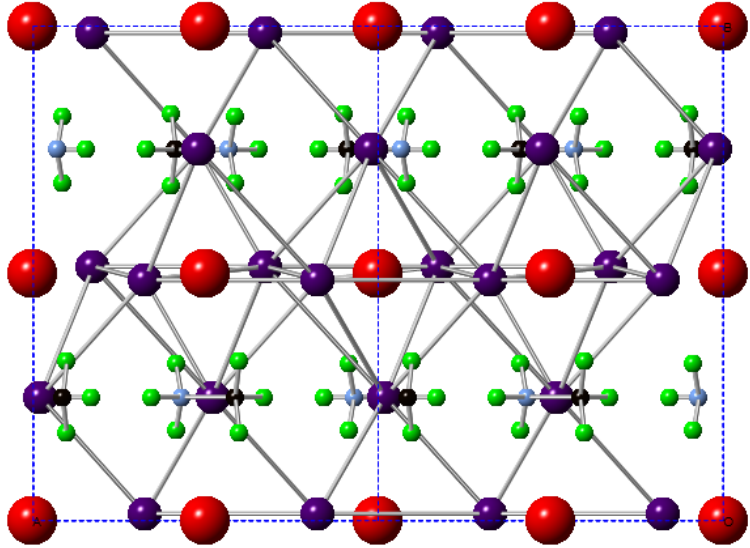

Fig. 1 showing 2x1x2 supercell for orthorhombic $CH_3NH_3PbI_3$. Red, purple, green, black, blue balls in the figure represent lead, iodine, hydrogen, carbon and nitrogen atoms respectively.

Contrary to the expectations that PBE would underestimate the bandgap [29,30], the bandgap value obtained is closer to experimental value of 1.6 eV. In addition, the reduced mass obtained from the PBE calculation was $0.18*m_e$, which is in fair agreement with experimental data of $0.15*m_e$. It indicates DFT can describe well the electronic properties of $MAPbI_3$. On the other hand, the bandgap obtained for Sn-based pervoksite is 0.80 eV with PBE while the experimental value is 1.2 eV, which is close to our HSE calculation of 1.15 eV. Due to such unbalanced description, the band-gap values are presented for all the proposed materials with various level of DFT theory as shown in Table. 1. It shows the bandgap values of obtained from various theories vary in a reasonable range. However, HSE+SO should be considered more accurate calculations compared to others. The lattice constants are shown with D2 correction. The lattice constants decrease due to inclusion of D2 corrections, which is in agreement with work by Feng et al. [16], discussed above. The effective mass values in x [100], y [010] and z [001] directions were calculated using PBE only as it is computationally expensive to calculate dense k-mesh energies using hybrid functionals. Some of the materials have very high effective masses compared to others, which is undesirable. Anisotropic effective masses were discovered in these materials, as discussed above, but anisotropic materials are also not suited for photovoltaics during uniform wafer fabrication.

Alkali metals iodide are generally in +1 oxidation state, hence they were excluded in the present search. Alkaline earth materials were selected as they have iodides in +2 state. Among them, magnesium shows promising feature of band-gap within the visible region, low effective mass values and reasonable absorption coefficient (Fig. 3a) compared to $MAPbI_3$ and $MASnI_3$. Different effective masses in different directions indicate the material could have anisotropic conductivity in different directions. In addition $MgI_2$ has crystallizes into P3m1 group which is similar to $PbI_2$, hence Mg is predicted as a potential replacement for Pb. Based on Goldscmidt's factor, the Ca based material should be more stable than the Mg, however Ca doesn't have band gap in visible region as shown in Table. 1. While alkaline earth metals are generally in +2 oxidation state, the transition elements can exist in multiple oxidation states, which is undesirable due to instability issue. Iodides of Cobalt (Co), Palladium (Pd), Zinc (Zn), Cdmium (Cd) are generally in +2 state, while Ti and V is prone to go to +4 state. However, $TiI_2$, $VI_2$, $CoI_2$ and $PdI_2$ have same space group as $PbI_2$. All of the transition metal organo halides studied here have the bandgap within visible region, however, based on the effective mass calculation Pd and Zn have lower effective masses. In addition, the Goldschmidt's factor for these materials is comparatively higher than the alkaline earth materials signifying their lesser stability. Among the transition metal series, Ti, V and Cd has relatively less Goldschmidt factor, but Ti has very high carrier effective mass. Based on the absorption coefficient (Fig. 2b) Zn, Cd and Co have relatively higher absorption compared to other transition metals discussed here. Furthermore, although the iodides of Ge and Si, which are in same periodic table group as Pb, have their iodides in +4 state, it was academically interesting to study their behavior using DFT. Both of the materials have bandgap in the visible region and they have reasonable better effective masses compared to other materials. Based on the absorption coefficient plot (Fig. 2c), these materials are predicted to have higher absorption compared to even Pb based compounds. For Si, there are no divalent iodides, however, for Ge the Goldschnmidt factor is not very high and based on its absorption behavior it is compelling to form this compound. Here, although the material is studied in orthorhombic group, it may lead to other phases such as tetragonal and cubic phase, but previously it has been that the band-gap and other relevant properties are not much affected after the phase change.

Table. 1 Band-gap, Goldschmidt's factor, lattice constant, electron and hole effective mass in different directions for various MAXI$_3$. PBE represents Perdew, Burke and Ernzerhof (PBE) functional, PBE+D2 is PBE with Grimme's D2 correction term. HSE stands for Heyd-Scuseria-Ernzerhof functional. +SO represents HSE along with spin orbit coupling. PBE and PBE+D2 are used for separate geometric optimization. However, for electronic calculations structures from PBE+D2 are used.

| X | Gap | | | | Gold | a | b | c | m$_h$ | | | m$_e$ | | |
|---|---|---|---|---|---|---|---|---|---|---|---|---|---|---|
|  | PBE | +D2 | HSE | +SO |  |  |  |  | [010] | [001] | [100] | [010] | [001] | [100] |
| Pb | 1.73 | 1.67 | 2.30 | 1.38 | 0.99 | 8.74 | 12.53 | 8.46 | 0.23 | 0.25 | 0.26 | 0.84 | 0.67 | 0.10 |
| Sn | 0.80 | 0.60 | 1.15 | 0.87 | - | 8.74 | 12.29 | 8.40 | 0.04 | 0.07 | 0.08 | 0.75 | 0.46 | 0.03 |
| Be | 2.70 | 2.53 | 3.54 | 3.39 | 1.28 | 8.49 | 12.39 | 7.40 | 0.56 | 1.62 | .6.79 | 0.74 | 0.47 | 0.47 |
| Mg | 1.47 | 1.76 | 2.68 | 2.49 | 1.16 | 8.39 | 11.78 | 7.93 | 0.56 | 1.24 | 1.83 | 0.36 | 0.28 | 0.28 |
| Ca | 3.77 | 3.82 | 4.99 | 4.73 | 1.05 | 8.67 | 12.33 | 8.32 | 0.47 | 2.12 | 5.61 | 0.53 | 1.30 | 0.94 |
| Sr | 3.84 | 3.94 | 5.02 | 4.72 | 1.04 | 8.82 | 12.74 | 8.75 | 0.69 | 2.37 | 11.96 | 0.39 | 0.38 | 0.31 |
| Ti | 1.52 | 0.54 | 1.91 | 2.26 | 1.10 | 8.27 | 11.36 | 7.84 | 137.2 | 0.83 | 0.68 | 31.2 | 0.55 | 0.51 |
| V | 0.10 | 0.09 | 0.64 | 2.72 | 1.13 | 8.15 | 11.28 | 7.71 | 48.7 | 0.9 | 0.8 | 4.12 | 3.97 | 0.31 |
| Co | 0.13 | 0.26 | 1.43 | 2.28 | 1.25 | 8.07 | 11.21 | 7.51 | 2.24 | 6.5 | 61.5 | 6.6 | 13.4 | 13.9 |
| Pd | 1.65 | 1.55 | 0.80 | 0.70 | 1.16 | 8.25 | 11.32 | 7.70 | 0.590 | 1.080 | 1.552 | 2.03 | 0.62 | 0.57 |
| Zn | 1.36 | 1.12 | 2.07 | 1.88 | 1.21 | 8.58 | 11.90 | 7.68 | 0.598 | 1.794 | 2.634 | 0.51 | 0.47 | 0.35 |
| Cd | 0.96 | 0.93 | 1.81 | 1.53 | 1.12 | 8.54 | 12.07 | 7.94 | 0.635 | 2.71 | 6.92 | 0.3 | 0.3 | 0.3 |
| Si | 0.08 | 0.04 | 0.22 | 0.56 | - | 8.43 | 11.55 | 7.98 | 0.029 | 0.049 | 0.1 | 0.03 | 0.81 | 0.52 |
| Ge | 0.84 | 0.72 | 1.25 | 1.09 | 1.15 | 8.48 | 11.84 | 8.06 | 0.08 | 0.11 | 0.13 | 0.78 | 0.61 | 0.05 |

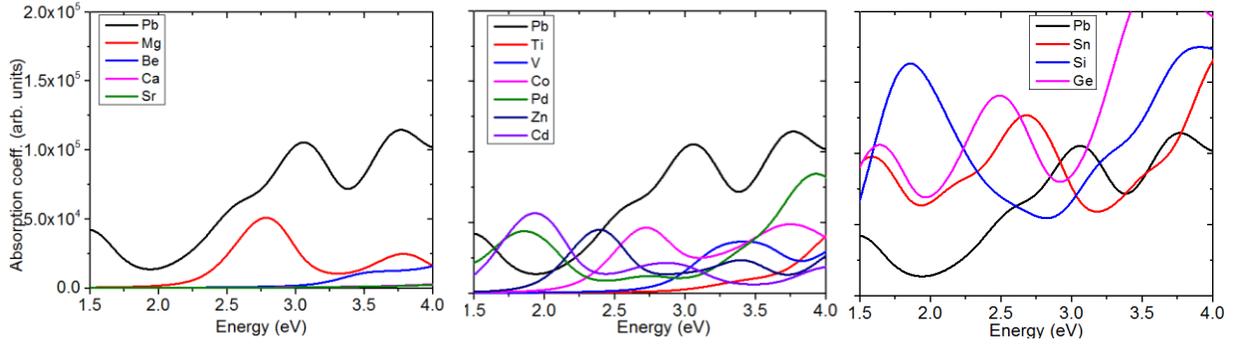

Fig. 2a          Fig. 2b          Fig. 2c

Fig. 2 Absorption coefficient obtained from HSE+SO calculation for the various materials discussed above. Among alkaline earth elements Mg, among transition elements Zn, Cd, Co and among group IVA Ge have comparatively high absorption among respective groups. Absorption of MAPbI$_3$ is also given for reference.

Further to the above-mentioned information, the absorption coefficient, the relative band alignment of the materials is also an important issue. Hence, the conduction band minima (CBM) and valence band maxima (VBM) data are compared for the materials as shown in Fig. 3. Among alkaline materials, Mg has closest resemblance to the VBM and CBM of MAPbI$_3$. Ca, Sr has very large differences of the CBM and VBM, which is also consistent with their bandgap. Hence they might not be suitable at all for solar cell applications. Among transition elements Zn, Cd and Co are better candidates for applications. Among group IVA, Ge and Si are possible replacements for Sn. However, as discussed earlier both of these materials are unstable due to their existence in +4 oxidation state.

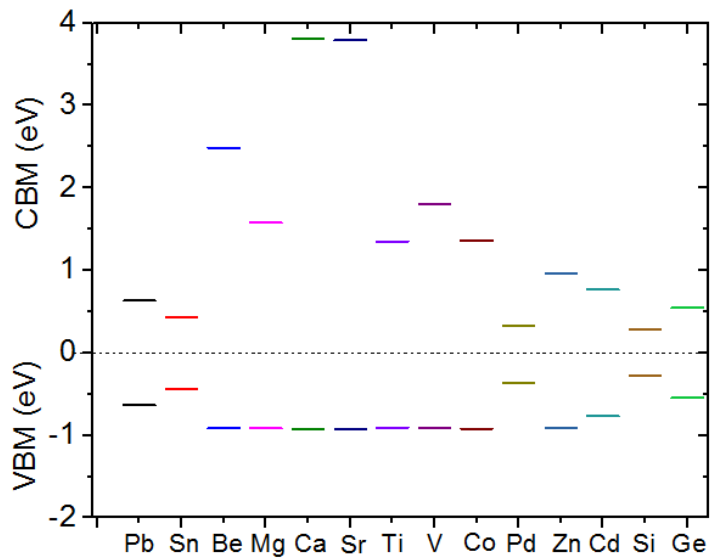

Fig. 2 Valence band maxima (VBM) and conduction band minima (CBM) positions for the proposed materials compared to MAPbI$_3$. Sn, Ti, Co, Zn, Cd, Ge are comparatively closer to MAPbI$_3$.

In summary, major problems in substituting Pb from $CH_3NH_3PbI_3$ are discussed and then elements from periodic table are identified that might be suitable for photovoltaic applications. From alkaline earth Mg, from transition elements Zn, Cd, Co and from group IVA Ge is predicted to be a replacement material apart from Sn, which has already been studied. Our results can guide experiments and save time during their search for lead-replacement material.

**Acknowledgement:** The author is thankful to Dr Susan B. Sinnott, Dr Jiangeng Xue and Dr Richard Hennig at University of Florida for helpful discussions.